\documentclass[12pt,tightenlines,eqsecnum,floats,aps,amsmath,amssymb,nofootinbib,prd]{revtex4}

\usepackage{setspace}
\usepackage{graphicx}

\usepackage{amsmath,amssymb,amsfonts,amsthm,mathrsfs}
\usepackage{enumerate}

\begin{document}

\title{Criticality in the collapse of spherically symmetric massless scalar fields in semi-classical loop quantum gravity}

\author{Florencia Benítez$^1$$^4$, Rodolfo Gambini$^1$,  Steven L. Liebling$^2$ and Jorge Pullin$^3$}
\affiliation{1. Instituto de F\'{\i}sica, Facultad de Ciencias, Igu\'a 4225, esq. Mataojo,
11400 Montevideo, Uruguay. \\
2. Long Island University, Brookville, New York 11548, USA.\\
3. Department of Physics and Astronomy, Louisiana State University,
Baton Rouge, LA 70803-4001, USA. \\
4. Instituto de F\'{\i}sica, Facultad de Ingenier\'{\i}a, Julio Herrera y Reissig 565,
11300 Montevideo, Uruguay.}

\begin{abstract}
In a recent paper we showed that the collapse to a black hole in one-parameter families of initial data for  massless, minimally coupled scalar fields in spherically symmetric semi-classical loop quantum gravity exhibited a universal mass scaling  similar to the one in classical general relativity. In particular, no evidence of a mass gap appeared as had been suggested by previous studies. The lack of a mass gap indicated the possible existence of a self-similar critical solution as in general relativity. Here we provide further evidence for its existence. Using an adaptive mesh refinement code, we show that  ``echoes'' arise as a result of the discrete self-similarity in space-time. We also show the existence of ``wiggles'' in the mass scaling relation, as in the classical theory. The results from the semi-classical theory agree well with those of classical general relativity unless one takes unrealistically large values for the polymerization parameter. 
\end{abstract}
\maketitle

\section{Introduction}

In 1993 Choptuik \cite{choptuik} studied numerically the collapse of a massless scalar field in spherically symmetric general relativity. Such a system has two possible final states: either the field disperses to infinity or forms a black hole. He concentrated on one-parameter families of initial data. Starting from values of the parameter for which no black hole forms, if one varies the parameter one eventually passes a``critical value'' for which it does. Three novel observations were made. On the one hand, no minimal black hole mass appears to exist. In other words,  one can create black holes as small as desired by fine tuning the initial parameter. Although this was somewhat expected since the problem does not have any characteristic mass scale, there was some debate about the existence of a gap before Choptuik's results. The second observation is that the mass of the resulting black hole 
depends on the distance in parameter-space from the critical value as a power law. The exponent of the power law is universal: it takes the same value for all families of initial data. In addition, Choptuik used sophisticated adaptive mesh refinement techniques to study the features of the solutions for values of the parameter very close to criticality. This study led to the third surprising behavior: as one approaches criticality the solution exhibited a discrete self-similarity in space-time. 

Let us put these results a bit more quantitatively. The universality in the exponent means that the mass of the black hole formed $m_\mathrm{BH}$ satisfies $m_\mathrm{BH}\approx\mathbb{C}\left|p-p^{\ast}\right|^{\gamma}$ where $p$ is the parameter characterizing the initial data and $p^{\ast}$ its critical value. This relationship, strictly speaking, holds in the limit $p\rightarrow p^{\ast}$. The universal exponent takes the value $\gamma\approx 0.37$ for all the initial data families studied within this model. The values of $p^{\ast}$ and $\mathbb(C)$ are family dependent.

Subsequently, Hod and Piran \cite{hodpiran} noted that there was a small correction to the above law
\begin{equation}
\ln\left(m_\mathrm{BH}\right)\approx\mathbb{\gamma\ln}\left|p-p^{\ast}\right|+c_{f}+\Psi[(\ln|p\text{\textminus}p^{*}|].\label{eq:1}
\end{equation}
with $\Psi[\ln|p-p^{*}|]$ taking the form of a ``wiggle'' of universal character.

The result for discrete self similarity says that if one considers a variable $Z(r,t)$ (representing say the scalar field or the metric components $g_{tt}$ or $g_{rr}$)  in the problem and writes it in terms of logarithmic coordinates $\rho,\tau$ (for the detailed relation to $r,t$ see \cite{choptuik}) one has that $Z(\rho-\Delta,\tau-\Delta)\sim Z(\rho,\tau)$ with $\Delta=3.4$ a universal constant independent of the initial data. Due to the self-similarity manifesting itself in the logarithmic coordinates, this ``echoing'' behavior is hard to see if one uses codes with a fixed mesh, as it would require a very small grid spacing throughout the mesh. It is better tackled using adaptive mesh refinement.

In a recent paper \cite{us} we have revisited these results using an approach to the semi-classical equations of loop quantum gravity. In reality, determining the true semi-classical equations is a hard problem
that requires identifying semiclassical states in the theory. Our current understanding of loop quantum gravity is not there yet, even for a simplified, spherically symmetric situation. In view of this we used a 
common approach to generate candidates for the semiclassical equations known as ``polymerization.'' In this approach, some of the variables of the theory that would be represented by holonomies (or point holonomies) in loop quantum gravity are replaced by a ``polymer'' version of them. For a generic variable $\phi$ this looks like $\phi\rightarrow \sin(k\phi)/k$ with $k$ the ``polymerization parameter.'' The justification for this comes from the fact that in loop quantum gravity variables like the connection are not well defined at the quantum level, but their holonomy (``their exponential'') is.

Therefore one replaces the variables by their exponentiated versions (the $\sin$ comes from the fact that one is interested in the real part of the exponentiated variable). The polymerization parameter would correspond in the case of connection type variables to the length of the loop along which one computes the holonomy. In the limit where that loop shrinks to a point, the exponentiated variable returns the original variable and one recovers the original classical theory. However, in loop quantum gravity one expects areas to be quantized and to have a minimum eigenvalue and therefore the length of a loop cannot be shrunk to zero, there will be a minimum value.  That minimum value would be the value of the polymerization parameter. The minimum eigenvalue of areas is related to the Planck length and therefore one expects polymerization parameters to be very small. Scalar fields also need to be polymerized to have a well defined Hilbert space compatible with diffeomorphism invariance \cite{thiemannscalar}.

Our study of the Choptuik phenomena using  the polymerized equations as a candidate for semiclassical loop quantum gravity revealed some surprising elements. On the one hand, in the slicing chosen by Choptuik, most of the variables to be polymerized disappear. The only one left is the scalar field itself. And therefore the polymer theory differs from general relativity coupled to a scalar field only in the terms involving the scalar field. Moreover, we observed that although the polymerization parameter now introduces a mass scale (the Planck mass), the scaling observed by Choptuik still remains: there is no observed minimal mass for the black holes one can form. This is in contrast to previous studies of the problem using polymerized metric variables \cite{husain} that found a mass gap. The critical exponent presented a dependence on the polymerization parameter, but it is very mild; for all practical values the results are indistinguishable from those of classical general relativity. Our study only used a fixed, uniform mesh code for simplicity, and so we were not able to observe the ``echoes'' of the discrete self-similarity, nor the ``wiggles'' observed by Hod and Piran \cite{hodpiran}.

In this paper we wish to explore the latter two issues, making use of a code with adaptive mesh refinement. We will observe that wiggles and echoes appear in the semi-classical theory and that their features differ little from those of classical general relativity unless one takes unrealistically large values of the polymerization parameter.

\section{Collapse in spherically symmetric semiclassical loop quantum gravity}

In \cite{us} we studied the collapse of a massless scalar field in spherical symmetry minimally coupled to gravity using semiclassical equations stemming from loop quantum gravity
 
Following the treatment of Choptuik \cite{choptuik}, we fix the radial coordinate to the usual Schwarzschild one so  $E^{x}=x^{2}$, where $E^{x}$ is the radial component of the triad.
This allows to eliminate the $K_x$ component from the problem by solving the diffeomorphism constraint. The ``polar'' condition that Choptuik chooses ($K_{x}^{x}=Tr(K)$) corresponds
in these variables to $K_{\varphi}=0$. With these choices, the gravitational part of the semi-classical equations reduces to classical general relativity and the only effect of the
semi-classical theory is in the polymerization of the scalar field, $\phi\rightarrow\frac{\sin\left(k\varphi\right)}{k}$, and its canonical momentum $P_{\phi}\rightarrow P_{\varphi}$, where $k$ is the polymerization parameter. We will take the resulting theory to be a candidate for a semi-classical approximation to the full quantum theory. This is based on the experience of many authors in the cosmological context, but is not guaranteed in ours. At the moment, no one knows how to do a full quantization of this system and derive a semi-classical approximation from it. The resulting equations are,
\begin{equation}
\frac{N'}{N}-\frac{\left(E^{\varphi}\right)'}{E^{\varphi}}+\frac{2}{x}-\frac{\left(E^{\varphi}\right)^{2}}{x^{3}}=0\label{eq:13}
\end{equation}

\begin{equation}
\frac{\left(E^{\varphi}\right)'}{E^{\varphi}}-\frac{3}{2x}+\frac{\left(E^{\varphi}\right)^{2}}{2x^{3}}-2\pi x\left(\frac{\left(P_{\varphi}\right)^{2}}{x^{4}}+\left(\varphi'\right)^{2}\cos^{2}\left(k\varphi\right)\right)=0\label{eq:14}
\end{equation}

\begin{equation}
\dot{\varphi}=\frac{4\pi N}{E^{\varphi}x}P_{\varphi}\label{eq:15}
\end{equation}

\begin{eqnarray}
\dot{P}_{\varphi}&=&
\frac{4\pi x^{2}}{E^{\varphi}}\left[
\left(\frac{3NE^{\varphi}-xN\left(E^{\varphi}\right)'+N'E^{\varphi}x}{E^{\varphi}}\right)\varphi'\cos^{2}\left(k\varphi\right)\right.\nonumber\\
&&\left.+xN\varphi''\cos^{2}\left(k\varphi\right)-xNk\left(\varphi'\right)^{2}\cos\left(k\varphi\right)\sin\left(k\varphi\right)\right]\label{eq:16}
\end{eqnarray}
where $N$ is the lapse, $E^\varphi$ the densitized triad in the $\varphi$ direction
and  $\varphi$
and $P_{\varphi}$ the polymerized scalar field and its conjugate momentum
and $k$ the polymerization parameter. More details of the equations can be see in \cite{us}.
We are using the same code we developed in that reference, based on Choptuik's original one, but in this paper we have turned on adaptive mesh refinement.

\section{Results for the mass scaling}
\begin{figure}[h]
\begin{centering}
\includegraphics[scale=0.25]{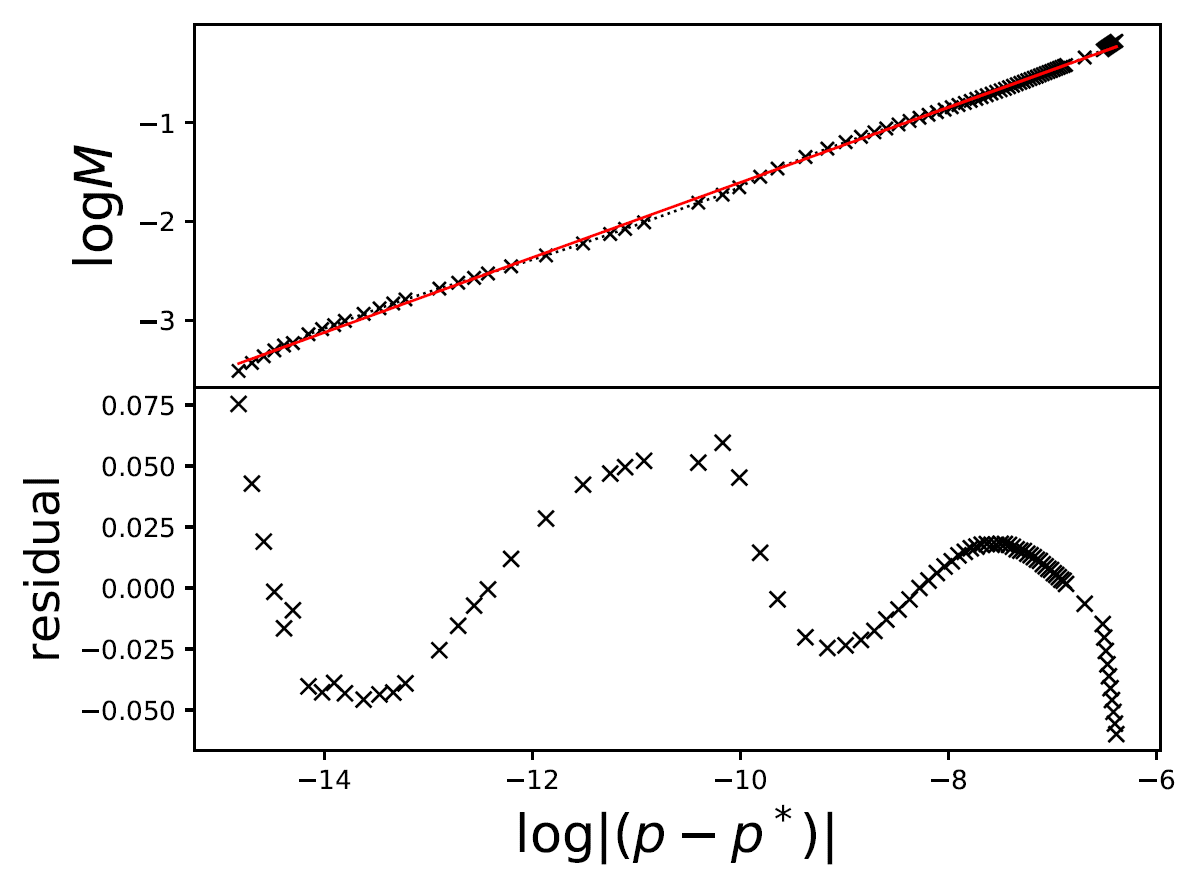}
\par\end{centering}
\caption{The mass scaling (upper panel) and the ``wiggles'' in the deviations from the linear fit (in log-log scale) first noted by Hod and Piran, for the case of general relativity (polymerized theory with $k=0$).
The crosses indicate the results of numerical evolutions and the red line is obtained as a least-squares fit to this data. The fit here has a slope $\gamma=0.37$, consistent with the value obtained by Choptuik. The range of the parameters considered here was chosen to maximize visibility of the ``wiggles,'' and does not necessarily give the best fit for the exponent, that is why we have less accuracy for it than in \cite{us}.
}
\label{fig1}
\end{figure}

Figure (\ref{fig1}) shows the power law of the mass as a function of the deviation of the parameter in the initial data in general relativity (polymerized theory with $k=0$). The initial data
consisted of a Gaussian profile in the scalar field parameterized by its amplitude. 
The mass scaling in the upper panel is quite linear modulated by a periodic wiggle, and the scaling is consistent with that observed by Choptuik.
The lower panel shows the deviations from the linear fit (in the log-log diagram), exhibiting the ``wiggles'' that Piran and Hod first noticed, consistent (within our accuracy)
with the expected periodicity $\varpi\approx 4.6$. 

\begin{figure}[h]
\begin{centering}
\includegraphics[scale=0.25]{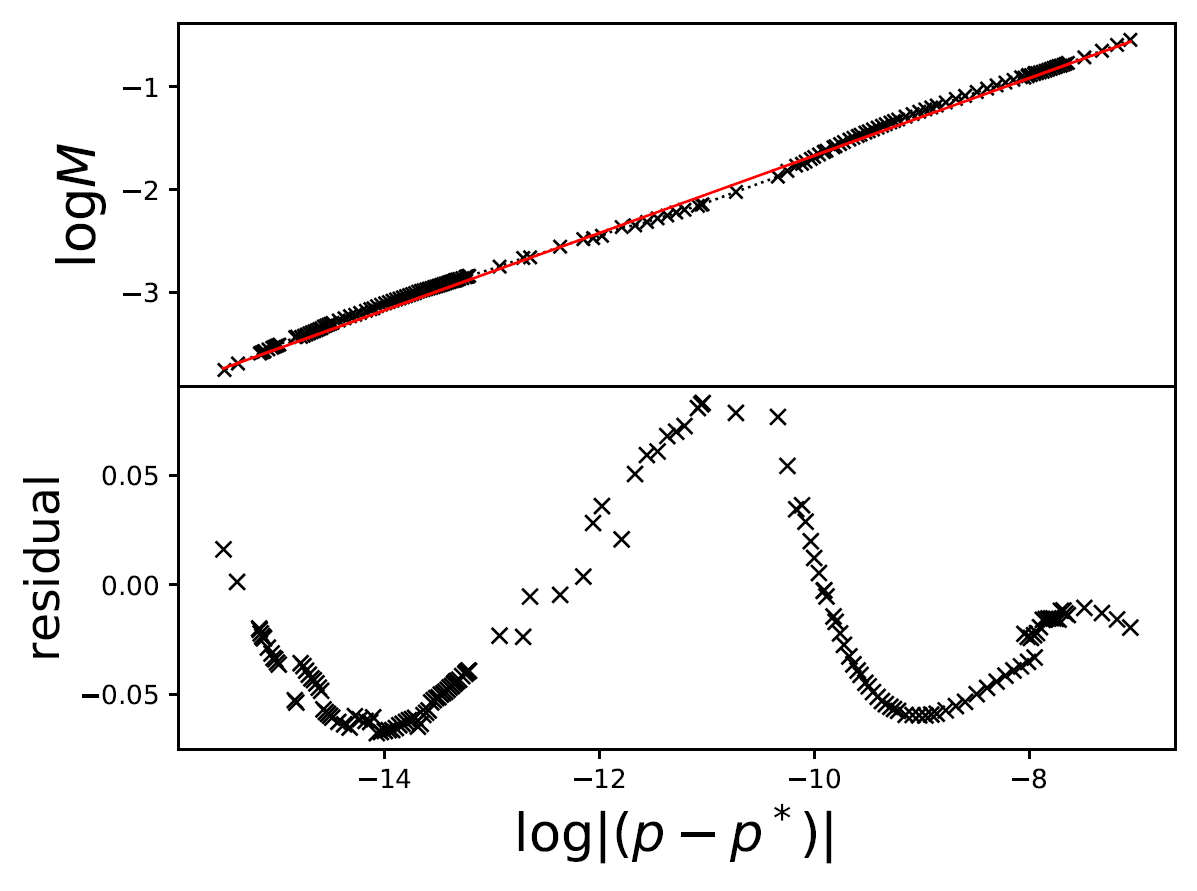}
\par\end{centering}
\caption{Mass scaling for $k=1$, similar to Fig. 1. The value $k=1$ is quite large, since $k$ is supposed to be a parameter of the order of Planck's length, which would be very small compared to the other length scales of the problem like the radius of the initial data shell being collapsed or the final black hole mass.}

\label{fig2}
\end{figure}

Figure (\ref{fig2}) shows the same results but now for the polymerized theory for $k=1$. It should be recalled that a realistic polymerization parameter value is determined by the Planck scale and therefore $k=1$ is unrealistically large. We choose this large value to make the effects more apparent. We see a mild dependence in the polymerization parameter for the mass scaling exponent ($0.38$ for $k=1$ vs. $0.37$ for $k=0$) and also on the frequency of the ``wiggles.'' Within numerical errors we cannot see a distinction between the $k=1$ and general relativity ($k=0$) cases.

\section{Discrete self-similarity}

\begin{figure}[h]
\begin{centering}
\includegraphics[scale=0.3]{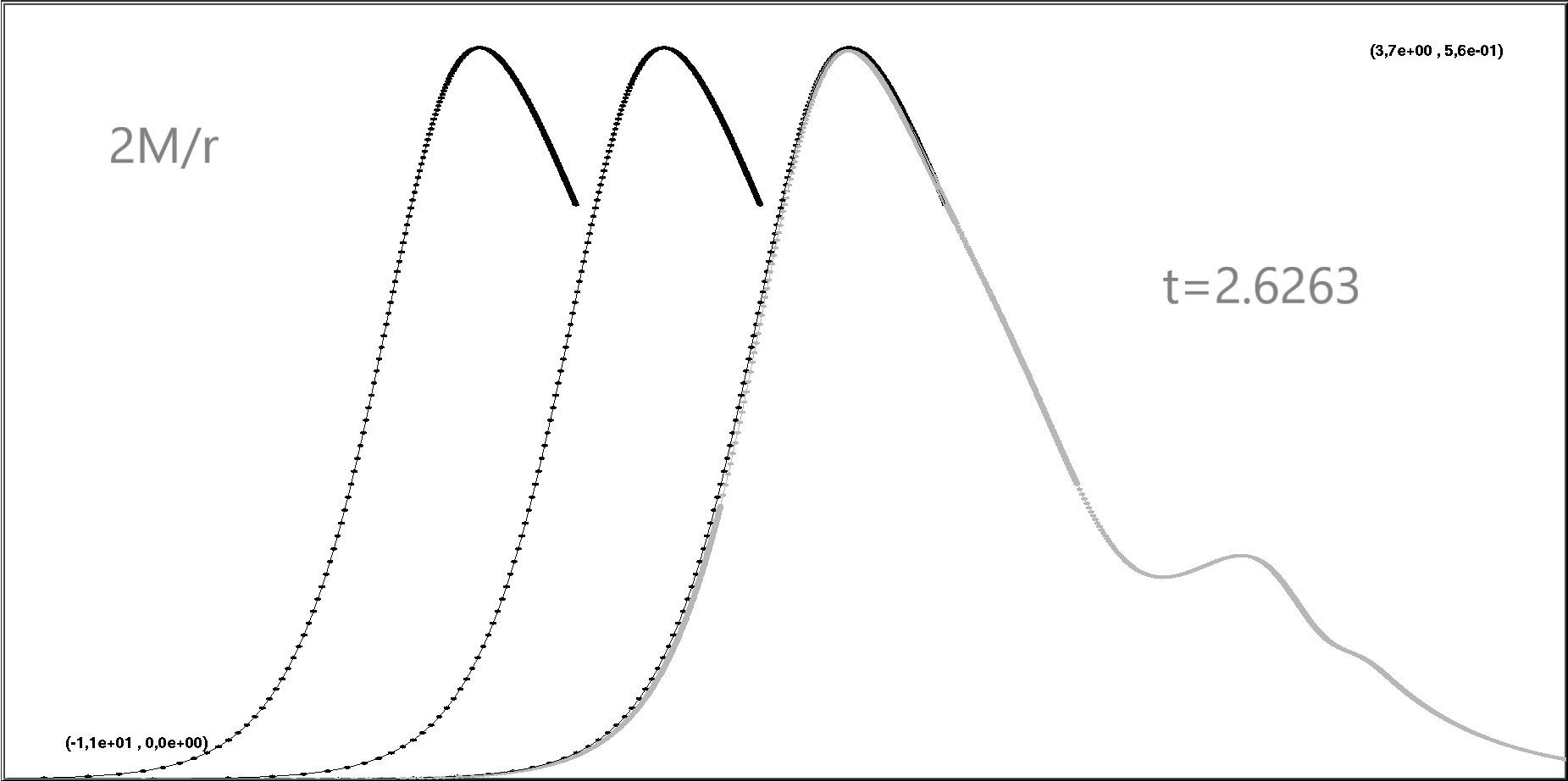}
\includegraphics[scale=0.3]{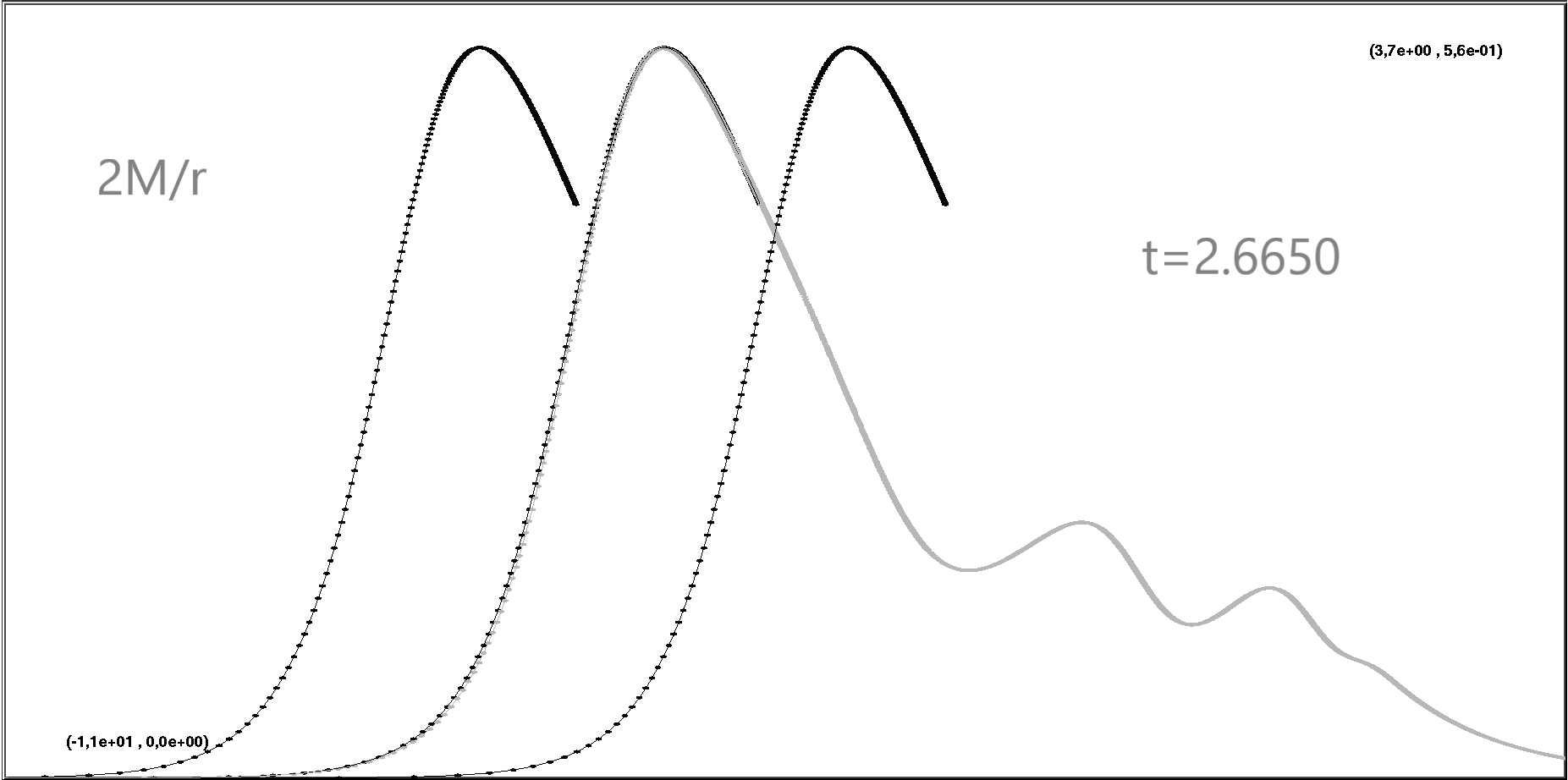}
\par\end{centering}
\caption{The curve $2M(r)/r$ plotted versus. $\ln(r)$ for the polymerized case $k=1$. To make the echoing apparent, the field at a late time ($t=2.6741$) is shown three times: The two bold curves to the right are the same curve translated an amount 1.7 and 3.4. The lighter curve to the right is the same field at two earlier times $t=2.6650$ (\textbf{top}), $2.6263$ (\textbf{bottom}). The agreement of these curves suggests the approach of the evolution to discrete self-similarity. From the curves we can determine that the period in the radial direction is $1.7$ and in time is $1.6$, in broad agreement with the ones observed by Choptuik of $3.4$ (since we chose a function that is positive definite, the proper period is twice the one observed with $2M(r)/r$ which ignores the sign of the scalar field). In the figure the radial axis ranges from $-11$ to $3.7$ and the vertical axis from $0$ to $0.56$.
}
\label{fig3}
\end{figure}

Choptuik noticed that his numerical evolutions very close to
criticality demonstrated a periodicity with an accompanying
change of scale. To demonstrate this behavior, we consider
a representative field in the critical regime at times periodic in log time.
Figure (\ref{fig3}) shows the function $2M(r)/r$ (where $M(r)$ is the mass aspect function) versus $\ln(r)$ for the polymerized case $k=1$. Displacing this curve in the logarithmic radial coordinate outwards an amount $3.4$ gives a profile that matches well this same field at an earlier time $3.2$ units before.
The agreement of these two curves suggests that the solution is repeating on a smaller scale consistent with discrete self-similarity\footnote{The period in the time direction has a larger numerical error because the periodicity is strictly speaking \cite{choptuik} defined in $\rho=\ln r$ and 
$\tau=\ln\left(T^*_0-T_0\right)$ where $T^*_0$ is the central proper time at which critical evolution stops, and its determination introduces further error.}. Because these observed periods (for log time and separately log radius) for $k=1$ are consistent with Choptuik's values, we do not detect a change in the critical solution due to the polymerization.


\section{Conclusions}

We have shown, using simulations of a massless scalar field coupled to the polymerized equations that may represent semi-classical loop quantum gravity, that the universality and scaling observed by Choptuik are still present. Also present are the ``wiggles'' in the exponent observed by Hod and Piran. The only effect of the polymerization is to alter slightly the values of the exponential scaling of the mass. We do not detect, within numerical accuracy, dependence of the period of the discrete self-similarity on the polymerization parameter. This provides robust evidence for the existence of a critical solution in the semi-classical regime  and that the solution appears to have the same periodicity of that of classical general relativity.

It should be emphasized that the semi-classical theory considered has not been derived from loop quantum gravity as at present no one knows how to build a quantum theory for this system. However, it offers an indication of what is possible for such a theory. It should also be emphasized that the semi-classical theory has a scalar field that is non-linear and may exhibit shocks. Such behavior is likely to occur close to the origin when one considers situations close to criticality, as large curvatures are expected there. We have not analyzed this phenomenon in detail and may study it in future publications. Even such an analysis should be taken with caution as in that region it is likely that any semi-classical theory may fail. It might be that a proper understanding of the Choptuik phenomena near the origin close to criticality will require a full quantization.

\section*{Acknowledgements}
We wish to thank Luis Lehner for discussions. 
This work was supported in part by Grants NSF-PHY-1903799, NSF-PHY-2011383, funds of the Hearne Institute for Theoretical Physics, CCT-LSU,  Pedeciba, Fondo Clemente Estable FCE\_1\_2019\_1\_155865.

\end{document}